\documentstyle[12pt,epsf]{article}
\begin{document}
%\tolerance=5000
\def\pp{{\, \mid \hskip -1.5mm =}}
\def\cL{{\cal L}}
\def\be{\begin{equation}}
\def\ee{\end{equation}}
\def\bea{\begin{eqnarray}}
\def\eea{\end{eqnarray}}
\def\tr{{\rm tr}\, }
\def\nn{\nonumber \\}
\def\e{{\rm e}}
\def\D{{D \hskip -3mm /\,}}

\ \hfill
\begin{minipage}{3.5cm}
NDA-FP-59 \\
April 1999 \\
\end{minipage}

\vfill

\begin{center}
{\Large\bf Can primordial wormholes be induced by GUTs at
the early Universe?}

\vfill

{\sc S. Nojiri}\footnote{
e-mail: nojiri@cc.nda.ac.jp}, 
{\sc O. Obregon$^{\clubsuit}$}\footnote{e-mail: 
octavio@ifug3.ugto.mx}, 
{\sc S.D. Odintsov$^{\spadesuit,\diamondsuit}$}\footnote{
e-mail: odintsov@mail.tomsknet.ru, odintsov@itp.uni-leipzig.de} 
and {\sc K.E. Osetrin$^{\spadesuit}$}\footnote{
e-mail: osetrin@tspu.edu.ru}

\vfill

{\sl Department of Mathematics and Physics \\
National Defence Academy, 
Hashirimizu Yokosuka 239, JAPAN}

\ 

{\sl $\clubsuit$
Instituto de Fisica de la Universidad de Guanajuato \\
P.O. Box E-143, 37150 Leon Gto., MEXICO}

\ 

{\sl $\spadesuit$ 
Tomsk State Pedagogical University, 634041 Tomsk, RUSSIA}

\ 

{\sl $\diamondsuit$ NTZ, University of Leipzig, Augustusplatz 10/11 \\
04109 Leipzig, GERMANY}

\

\vfill

{\bf abstract} 

\end{center}

Using large $N$, 4d anomaly induced one-loop effective action for
conformally invariant matter (typical GUT multiplet) we study the
possibility to induce the primordial spherically symmetric wormholes 
at the early Universe. The corresponding effective equations are 
obtained in two different coordinate frames. The numerical 
investigation of these equations is done for matter content 
corresponding to ${\cal N}=4$ $SU(N)$ super Yang-Mills theory. 
For some choice of initial conditions, the induced wormhole solution 
with increasing throat radius and increasing
red-shift function is found.

\newpage

The wormholes are puzzling topological objects (kind of
handles of topological origin) which attract much attention
in General Relativity for years. It is not strange as they may 
be considered as bridges joining two different Universes or two 
separate regions of the same Universe (for an introduction, see 
refs.\cite{12}). There are many speculations related with 
hypotetical effects which may be expected near wormholes.

Moreover, it is known that wormholes usually cannot occur as
classical solutions of gravity (with matter) due to violations
of energy conditions \cite{12}. Nevertheless, one can expect
that primordial wormholes may present at the very early Universe
where quantum effects play an essential role (or yet in
fundamental M-theory, see,for example, \cite{13}).

Indeed as it was shown in ref.\cite{15} (see also \cite{16})
using quantum stress tensor for conformal scalar on 
spherically symmetric space and in ref.\cite{NOOO} using
one-loop effective action in large $N$ and $s$-wave approximation 
for minimal scalar there may exist semiclassical quantum solution 
corresponding to a Lorentzian wormhole connecting two asymptotically 
flat regions of the Universe.
The corresponding spherically symmetric wormhole 
solution has been found numerically in both approximations
\cite{15,NOOO} as well as analytically \cite{NOOO}. That 
shows the principal possibility of inducing  primordial 
wormholes at the early Universe (in its quantum regime). However 
above discussion has been limited strictly to scalar matter. And 
what happens if other types of matter are included? Hence, the very 
natural question is: Can primordial wormholes be induced from GUTs 
at the early Universe?

In the present Letter we try to answer this question for spherically 
symmetric wormholes.
We use 4d conformal anomaly induced effective action in one-loop
and large $N$ approximation so that one can neglect quantum
gravitational contributions. The effective equations for an arbitrary 
massless GUT
containing conformal scalars, spinors and vectors are explicitly
obtained for two forms of spherically symmetric background.
Numerical solution of these equations is presented for ${\cal N}=4$
super Yang-Mills theory. It is observed that it depends on the
choice of the initial conditions. The initial conditions admitting
inducing of primordial wormholes are also presented.

We first derive the effective action for 
conformally invariant matter (for a general review of 
effective action in curved space, see\cite{BOS}). 
Let us start from Einstein gravity with $N_0$ 
conformal scalars $\chi_i$, $N_1$ vectors $A_\mu$ and
$N_{1/2}$ Dirac spinors $\psi_i$
\bea
\label{OI}
S&=&-{1 \over 16\pi G}\int d^4x \sqrt{-g_{(4)}}
\left\{R^{(4)} -2\Lambda\right\}
+ \int d^4x \sqrt{-g_{(4)}}\left\{{1 \over 2}\sum_{i=1}^{N_0}
\left(g_{(4)}^{\alpha\beta}\partial_\alpha\chi_i
\partial_\beta\chi_i \right.\right. \nn
&& \left.\left.+ {1 \over 6}R^{(4)}\chi_i^2 \right)
-{1 \over 4}\sum_{j=1}^{N_1}F_{j\,\mu\nu}F_j^{\mu\nu}
+\sum_{k=1}^{N_{1/2}}\bar\psi_k\D\psi_k \right\}\ .
\eea
The above matter content is typical for asymptotically free
GUT at high energies \cite{ BOS} as interaction terms
and masses at strong curvature
 are negligible due to asymptotic freedom. For finite GUTs
or asymptotically non -free GUTs
one should not include masses if consider only conformally
invariant theories and interaction plays no role as anyway we consider
purely gravitational background.

The convenient choice for the spherically symmetric
space-time is the following:
\be
\label{OIV}
ds^2=f(\phi)\left[
f^{-1}(\phi)g_{\mu\nu}dx^\mu dx^\nu + r_0^2 d\Omega\right]\ .
\ee
%\be
%\label{OII}
%ds^2=g_{\mu\nu}dx^\mu dx^\nu + f(\phi) r_0^2 d\Omega
%\ee
where $\mu,\nu=0,1$, $g_{\mu\nu}$ and $f(\phi)$ depend only
from $x^1$ and $r_0^2$ is non-essential constant.

Let us start the calculation of matter effective action
 on the background (\ref{OIV})
closely following to ref.\,\cite{NO}(see also \cite{NOa}).
In the calculation of effective action, we present
effective action as :
$\Gamma=\Gamma_{ind}+ \Gamma[1, g^{(4)}_{\mu\nu}]$
where $\Gamma_{ind}=\Gamma[f, g^{(4)}_{\mu\nu}]
- \Gamma[1, g^{(4)}_{\mu\nu}]$ is conformal anomaly
induced action which is quite well-known \cite{R},
$g^{(4)}_{\mu\nu}$ is metric (\ref{OIV}) without
multiplier in front of it, i.e., $g^{(4)}_{\mu\nu}$
corresponds to
\be
\label{OVI}
ds^2=\left[
\tilde g_{\mu\nu}dx^\mu dx^\nu + r_0^2 d\Omega\right]\ ,\ \ \
\tilde g_{\mu\nu}\equiv f^{-1}(\phi)g_{\mu\nu} \ .
\ee

The conformal anomaly for above matter is well-known
\be
\label{OVII}
T=b\left(F+{2 \over 3}\Box R\right) + b' R_{GB} + b''\Box R
\ee
where $b={(N_0 +6N_{1/2}+12N_1)\over 120(4\pi)^2}$,
$b'=-{(N_0+11N_{1/2}+62N_1) \over 360(4\pi)^2}$, $b''=0$ but
in principle, $b''$  may be changed by the finite
renormalization of local counterterm in gravitational
effective action, $F$
is the square of Weyl tensor, $R_{GB}$ is
Gauss-Bonnet invariant.

Conformal anomaly induced effective action $\Gamma_{ind}$
may be written as follows \cite{R}:
\bea
\label{OVIII}
W&=&b\int d^4x \sqrt{-g} F\sigma
+b'\int d^4x \sqrt{-g} \Bigl\{\sigma\left[
2\Box^2 + 4 R^{\mu\nu}\nabla_\mu\nabla_\nu \right. \nn
&& \left. - {4 \over 3}R\Box + {2 \over 3}(\nabla^\mu R)\nabla_\mu
\right]\sigma + \left(R_{GB}-{2 \over 3}\Box R\right)\sigma \Bigr\} \nn
&& -{1 \over 12}\left(b'' + {2 \over 3}(b + b')\right)
\int d^4x \sqrt{-g}\left[\left\{R - 6 \Box \sigma
- 6(\nabla \sigma)(\nabla \sigma)\right\}^2 - R^2 \right]
\eea
where $\sigma={1 \over 2}\ln f(\phi)$, and
$\sigma$-independent terms are dropped. All 4-dimensional
quantities (curvatures, covariant derivatives) in
Eq.(\ref{OVIII}) should be calculated on the metric
(\ref{OVI}).
The calculation of $\Gamma[1, g_{\mu\nu}^{(4)}]$ is done in ref.
\cite{NO} in leading order as follows:
\be
\label{SDW}
\Gamma[1, g_{\mu\nu}^{(4)}]=\int d^4x
\sqrt{-g}\left\{\left[{b}F+  {b'}R_{GB}
+ { 2b\over 3}\Box R\right]\ln { R \over \mu^2}\right\}
+{\cal O}(R^3)
\ee
where $\mu$ is mass-dimensional constant parameter, all
the quantities are calculated on the background (\ref{OVI}).
The condition of application of above expansion is
$|R|<R^2$ (curvature is nearly constant), in this case
we may limit by only few first terms.

Let us solve the equations of motion obtained from
the above effective Lagrangians $S+\Gamma$.
In the following, we use $\tilde g_{\mu\nu}$ and $\sigma$
as a set of independent variables and we write
$\tilde g_{\mu\nu}$ as $g_{\mu\nu}$ if there is no
confusion.

$\Gamma_{ind}$ ($W$ in Eq.(\ref{OVIII})) is rewritten after the reduction
to 2 dimensions as follows:
\bea
\label{Gindrd}
{\Gamma_{ind} \over 4\pi}&=&
{b r_0^2 \over 3}\int d^2x\sqrt{-g}\left(
\left(R^{(2)} + R_\Omega\right)^2 + {2 \over 3}R_\Omega R^{(2)}
+ {1 \over 3}R_{\Omega}^2 \right) \sigma \nn
&& + b' r_0^2 \int d^2x\sqrt{-g}\left\{\sigma \left(
2\Box^2 + 4R^{(2)\mu\nu}\nabla_\mu \nabla_\nu
-{4 \over 3}(R^{(2)} + R_\Omega)\Box \right.\right. \nn
&& \left.
+ {2 \over 3}(\nabla^\mu R^{(2)})\nabla_\mu\right)
\sigma \left. + \left( 2R_\Omega R^{(2)} - {2 \over 3}\Box R^{(2)}
\right)\sigma \right\} \nn
&& -{1 \over 12}\left\{ b'' + {2 \over 3}(b+b')\right\}r_0^2
\int d^2x\sqrt{-g} \nn
&& \times \left\{ \left( R^{(2)} + R_\Omega - 6 \Box\sigma
- 6\nabla^\mu\sigma\nabla_\mu\sigma \right)^2
- \left( R^{(2)} + R_\Omega \right)^2 \right\}
\eea
Here $R_\Omega={2 \over r_0^2}$ is scalar curvature of $S^2$ with
the unit radius.

Let us derive the effective equations of motion.
In the following, we work in  the conformal gauge :
$g_{\pm\mp}=-{1 \over 2}\e^{2\rho}\ ,\ \ \ g_{\pm\pm}=0$
after considering the variation of the effective action
$\Gamma+S$ with respect to $g_{\mu\nu}$ and $\sigma$.
Note that the tensor $g_{\mu\nu}$ under consideration is
the product of the
original metric tensor and the $\sigma$-function
$\e^{-2\sigma}$, the equations given by the variations over
$g_{\mu\nu}$ are the combinations of the equations given by the
variation over the original metric and $\sigma$-equation.

It often happens that we can drop the terms linear in $\sigma$
in (\ref{Gindrd}). In particular, one can redefine the corresponding
source term as it is in the case of IR sector of 4D QG \cite{AMO}.
In the following, we only consider this case.
Then the variations of $S+\Gamma_{ind}+\Gamma[1, g^{(4)}_{\mu\nu}]$
with respect to $g^{\pm\pm}$, $\rho$ and $\sigma$ are given by
(see also \cite{NO})
\bea
\label{cons1}
0&=&-{ r_0^2 \over 48\pi G}\e^{2\rho+2\sigma}\left[(\partial_r\sigma)^2
-\partial_r^2\sigma+2\partial_r\sigma\partial_r\rho\right]
+ {b' r_0^2 \over 16}\left[ -8\e^{2\rho}\partial_r \sigma \partial_r
\left(\e^{-2\rho}\partial_r^2 \sigma \right) \right. \nn
&& \left. +8\sigma\partial_r^2\sigma \partial_r^2 \rho
+ {8 \over 3}\e^{2\rho}\partial_r\sigma \partial_r\left\{R_4
\sigma \right\} + {32 \over 3}\e^{2\rho}\sigma
\partial_r R_4\partial_r \sigma  \right] \nn
&& -\left\{b''+{2 \over 3}(b+b')\right\} {r_0^2 \over 16}\left[
16\e^{2\rho}\partial_r\sigma \partial_r R_4
+ 4 (\partial_r \sigma)^2\partial_r^2 \rho \right. \nn
&& \left. - 12 \e^{2\rho}\partial_r\sigma \partial_r
\left\{\e^{-2\rho}\left(\partial_r^2\sigma
+ \left(\partial_r \sigma\right)^2\right)\right\}
+12 \left(\partial_r^2\sigma
+ \left(\partial_r \sigma\right)^2 \right)(\partial_r\sigma)^2 \right] \nn
&& - {r_0^2 \over 16}
\left\{ {1 \over 2}\partial_r^2\rho - 2(\partial_r\rho)^2
-{1 \over 4} \partial_r^2 + {3 \over 2} \partial_r\rho \partial_r
\right\} \nn
&& \times \left[ {16 \over 3}b' \left(-\sigma\partial_r^2 \sigma
+ \left(\partial_r\sigma\right)^2\right)
-\left\{b''+{2 \over 3}(b+b')\right\}
\left(\partial_r^2\sigma + \left(\partial_r\sigma\right)^2 \right)
\right]\nn
&& + r_0^2
\left[-{1 \over 12}b \e^{2\rho}\partial_r\left\{\ln \left(
{R_4 \over \mu^2 } \right)\right\} \partial_r R_4 \right. \nn
&& - \left\{{1 \over 2}\partial_r^2\rho - 2(\partial_r\rho)^2
-{1 \over 4} \partial_r^2
+ {3 \over 2} \partial_r\rho \partial_r \right\} \nn
&& \times \left\{{2 \over 3}b\partial_r^2\rho \ln \left(
{R_4 \over \mu^2 } \right) \right. + {b \over 4}
\partial_r^2\left\{\ln \left({R_4 \over\mu^2}\right)\right\} \\
&& \left. \left. + \left\{b\left({2 \over 3}\e^{-2\rho}(\partial_r^2\rho)^2
+ {2\e^{2\rho} \over 3 r_0^2} + {1 \over 3}\partial_r^2R_4
\right) +{4 \over r_0^2}\left({b \over 3}+b'\right) \partial_r^2\rho\right\}
{1 \over R_4}\right\} \right] \ , \nn
\label{R4}
R_4&\equiv& -2\e^{-2\rho}\partial_r^2\rho + {2 \over r_0^2}\ ,\\
\label{rhov}
0&=&-{ r_0^2 \over 16\pi G} \left[-\partial_r^2\e^{2\sigma}
+ {4 \over r_0^2}\e^{2\rho+2\sigma}\right]
+b' r_0^2 \left\{ -2(\partial_r^2\sigma)^2\e^{-2\rho} \right.\nn
&&-{8 \over 3}\e^{-2\rho} \partial_r^2\rho (\sigma\partial_r^2\sigma)
+ {4 \over 3}\partial_r^2\left(\e^{-2\rho}
\sigma\partial_r^2\sigma\right) \nn
&& \left. - {1 \over 3}\left\{-2\left(\partial_r\sigma\right)^2
\e^{-2\rho}\partial_r^2\rho
+ \partial_r^2\left(\left(\partial_r\sigma\right)^2
\e^{-2\rho}\right)\right\}\right\} \nn
&& -\left\{b''+ {2 \over 3}(b+b')\right\} r_0^2\left[
\partial_r^2 \left\{ \e^{-2\rho} \left(\partial_r^2\sigma
+\left(\partial_r\sigma\right)^2\right)\right\} \right. \nn
&& \left. - 3\e^{-2\rho}\left(\partial_r^2 \sigma
+\left(\partial_r\sigma\right)^2\right)^2\right] \nn
&& +r_0^2\left[ -{4 \over 3}b\e^{-2\rho}\left(\partial_r^2
\rho \right)^2 \ln \left( {R_4 \over \mu^2 } \right) \right.
+{4 \over 3}b\partial_r^2\left\{\e^{-2\rho}\partial_r^2
\rho \ln \left( {R_4 \over \mu^2 } \right)\right\} \nn
&& + {4b\e^{2\rho} \over 3 r_0^2} \ln \left({R_4 \over \mu^2 } \right)
-{4 \over r_0^2}\left({b \over 3}+b'\right)\partial_r^2
\ln \left({R_4 \over \mu^2 } \right) \\
&& +{4 \over 3}b\e^{-2\rho}\partial_r^2\rho \partial_r^2
\left\{\ln \left( {R_4 \over \mu^2 } \right)\right\}
-{4 \over 3}b\partial_r^2\left\{\e^{-2\rho} \partial_r^2
\left\{\ln \left( {R_4 \over \mu^2 } \right)\right\}\right\} \nn
&& + {4\e^{-2\rho}\partial_r^2\rho \over R_4 }
\left\{{2 \over 3}b\e^{-2\rho}\left(\partial_r^2\rho\right)^2
+ {2b\e^{2\rho} \over 3 r_0^2}
+ {1 \over 3}b\partial_r^2 R_4
- {4 \over r_0^2}\left({b \over 3}+b'\right)\partial_r^2
\rho \right\} \nn
&& + \partial_r^2\left\{{2\e^{-2\rho}\over R_4 }
\left\{{2 \over 3}b\e^{-2\rho}
(\partial_r^2\rho)^2+ {2b\e^{2\rho} \over 3r_0^2}
\left. + {1 \over 3}b\partial_r^2 R_4 - {4 \over r_0^2}
\left({b \over 3}+b'\right)
\partial_r^2\rho \right\}\right\}\right] \ , \nn
\label{sigmav}
0&=& -{ r_0^2 \over 16\pi G}
\Biggl[-2\e^{2\sigma}\left\{3\partial_r^2 \sigma
+3\left(\partial_r\sigma\right)^2 + \partial_r^2 \rho \right\}
+ {4 \over r_0^2}\e^{2\rho+2\sigma} \Biggr] \nn
&& + b' r_0^2\left[ 2 \partial_r^2 (\e^{-2\rho}\partial_r^2 \sigma)
+{4 \over 3}\left(\e^{-2\rho}\partial_r^2 \rho
\partial_r^2 \sigma + \partial_r^2(\sigma\e^{-2\rho}
\partial_r^2 \rho)\right) \right. \nn
&& \left. - {8 \over 3 r_0^2}\partial_r^2 \sigma
- {1 \over 6}\left\{\partial_r R_4\partial_r \sigma
+\partial_r R_4\partial_r \sigma\right\} \right]\nn
&& -\left\{b''+ {2 \over 3}(b+b')\right\} r_0^2\left[
-{1 \over 2}\partial_r^2 R_4
+{1 \over 2}\left\{\partial_r R_4\partial_r \sigma +\partial_r R_4
\partial_r\sigma\right\} \right. \nn
&& \left. + 3 \partial_r^2 \left\{ \e^{-2\rho}
\left(\partial_r^2\sigma +\left(\partial_r\sigma\right)^2\right)\right\}
-6 \partial_r\left\{ \e^{-2\rho}\partial_r\sigma\left(\partial_r^2
\sigma + \left(\partial_r\sigma\right)^2\right)\right\}  \right]
\eea
Note that now the real 4 dimensional metric is given by
\be
\label{metric}
ds^2=-\e^{2\sigma + 2\rho}dx^+dx^- + r_0^2\e^{2\sigma}d\Omega^2\ .
\ee
Let us make the following change: $\partial_r$ etc. as
\be
\label{chdr}
\partial_r = \e^{\rho+\sigma}\partial_l\ ,\ \
\partial_r^2 = \e^{2\rho+2\sigma}\left(\partial_l^2
+ \partial_l\rho\partial_l + \partial_l\sigma\partial_l \right)\ .
\ee
 Using the new radial coordinate $l$,
the metric (\ref{metric}) is rewritten as follows:
\bea
\label{lmetric}
ds^2&=& - f(l)dt^2 + dl^2 + r(l)^2 d\Omega^2 \nn
f(l)&=&\e^{2\sigma+2\rho} \nn
r(l)&=&r_0\e^\sigma\ .
\eea
Here $f(l)$ is called a redshift function and $r(l)$ is a shape
function. If $f(l)$ and $r(l)$ are smooth positive-definite
functions, which satisfy
the conditions:
\bea
\label{cond}
&f(l)\rightarrow 1,\  r(l)\rightarrow l \ & \mbox{when}\
|l|\rightarrow \infty \nn
&f(l),\ r(l) \rightarrow \ \mbox{finite} & \mbox{when}\
|l|\rightarrow 0\ ,
\eea
the metric expresses the wormhole which connects two asymptotically
flat universes.

 For the choice of the metric (\ref{lmetric}),
Eqs.(\ref{cons1}), (\ref{R4}), (\ref{rhov}) and (\ref{sigmav})
are rewritten as follows:
%%%%%
\bea
\label{cons}
0&=&-{ r_0^2 \over 48\pi G}\e^{4\rho+4\sigma}\left[
-\partial_l^2\sigma + \partial_l\sigma\partial_l\rho\right] \nn
&& + {b' r_0^2 \e^{4\rho+4\sigma} \over 16}\left[
-8\e^{-2\sigma}\partial_l \sigma \partial_l \left(\e^{2\sigma}
\left(\partial_l^2 \sigma + \partial_l\rho\partial_l \sigma
+ (\partial_l\sigma)^2 \right)\right) \right. \nn
&& +8\sigma\left(\partial_l^2 \sigma
+ \partial_l\rho\partial_l \sigma+ (\partial_l\sigma)^2 \right)
\left(\partial_l^2 \rho + \left(\partial_l\rho\right)^2
+ \partial_l\sigma\partial_l\rho\right) \nn
&& \left. + {8 \over 3}\e^{-2\sigma}\partial_l\sigma \partial_l\left\{R_4
\sigma \right\} + {32 \over 3}\e^{-2\sigma}\sigma
\partial_l R_4\partial_l \sigma  \right] \nn
&& -\left\{b''+{2 \over 3}(b+b')\right\} {r_0^2 \e^{4\rho+4\sigma}
\over 16}\left[16\e^{-2\sigma}\partial_l\sigma \partial_l R_4
+ 4 (\partial_l \sigma)^2\left(\partial_l^2 \rho
+ \left(\partial_l\rho\right)^2 + \partial_l\sigma\partial_l\rho
\right) \right. \nn
&& - 12 \e^{-2\sigma}\partial_l\sigma \partial_l
\left(\e^{2\sigma}\left(\partial_l^2\sigma + \partial_l\rho\partial_l \sigma
+ 2 \left(\partial_l \sigma\right)^2\right)\right) \nn
&& \left. +12 \left(\partial_l^2\sigma + \partial_l\rho\partial_l \sigma
+ 2 \left(\partial_l \sigma\right)^2 \right)(\partial_l\sigma)^2 \right] \nn
&& - {r_0^2 \e^{2\rho+2\sigma} \over 16}
\left\{{1 \over 2} \partial_l^2\rho - {3 \over 2}(\partial_l\rho)^2
+ {1 \over 2} \partial_l\sigma\partial_l\rho -{1 \over 4} \partial_l^2
+ {5 \over 4} \partial_l\rho \partial_l - {1 \over 4}
\partial_l\sigma\partial_l \right\} \nn
&& \times \e^{2\rho+2\sigma} \left[ {16 \over 3}b'
\left(-\sigma\left(\partial_l^2 \sigma
+ \partial_l\rho\partial_l \sigma + (\partial_l\sigma)^2 \right)
+ \left(\partial_l\sigma\right)^2\right) \right. \nn
&& \left. -\left\{b''+{2 \over 3}(b+b')\right\}
\left(\partial_l^2\sigma + \partial_l\rho\partial_l \sigma
+ 2 \left(\partial_l\sigma\right)^2 \right) \right] \nn
&& + r_0^2 \left[-{1 \over 12}b \e^{4\rho + 2\sigma}
\partial_l\left\{\ln \left(
{R_4 \over \mu^2 } \right)\right\} \partial_l R_4 \right. \\
&& - \e^{2\rho + 2\sigma} \left\{{1 \over 2} \partial_l^2\rho
- {3 \over 2}(\partial_l\rho)^2
+ {1 \over 2}\partial_l\sigma\partial_l\rho -{1 \over 4} \partial_l^2
+ {5 \over 4} \partial_l\rho \partial_l
- {1 \over 4} \partial_l\sigma\partial_l \right\} \nn
&& \times \e^{2\rho + 2\sigma} \left\{{2 \over 3}b
\left(\partial_l^2 \rho + \left(\partial_l\rho\right)^2
+ \partial_l\sigma\partial_l\rho \right) \ln \left(
{R_4 \over \mu^2 } \right) \right. \nn
&& + {b \over 4}\left(\partial_l^2  + \partial_l\rho \partial_l
+ \partial_l\sigma\partial_l \right)
\left\{\ln \left({R_4 \over\mu^2}\right)\right\} \nn
&& + \left\{ b\left({2 \over 3}\e^{-2\rho}
\left(\partial_l^2 \rho + \left(\partial_l\rho\right)^2
+ \partial_l\sigma\partial_l\rho \right)^2
+ {2\e^{-2\sigma} \over 3 r_0^2} \right.\right.\nn
&&
\left.+ {1 \over 3}
\left(\partial_l^2 + \partial_l\rho \partial_l
+ \partial_l\sigma\partial_l \right) R_4 \right) \nn
&& \left. \left. \left.
+{4 \over r_0^2}\left({b \over 3}+b'\right) \left( \partial_l^2\rho
+ \left(\partial_l\rho\right)^2 + \partial_l\sigma\partial_l\rho \right)
\right\}{1 \over R_4}\right\} \right] \ , \nn
\label{R4b}
R_4&\equiv& -2\e^{-2\sigma}\left(\partial_l^2 \rho
+ \left(\partial_l\rho\right)^2
+ \partial_l\sigma\partial_l\rho \right) + {2 \over r_0^2}\ ,\\
\label{rho}
0&=&-{ r_0^2 \e^{2\rho} \over 16\pi G} \left[
-\e^{2\sigma}\left(\partial_l^2 + \partial_l\rho \partial_l
+ \partial_l\sigma\partial_l \right) \e^{2\sigma}
+ {4 \over r_0^2}\e^{2\sigma}\right] \nn
&& +b' r_0^2 \e^{2\rho + 4\sigma} \left\{ -2 \left(\partial_l^2 \sigma
+ \partial_l\rho\partial_l \sigma
+ (\partial_l\sigma)^2 \right)^2 \right. \nn
&& -{8 \over 3}\left(\partial_l^2 \rho
+ \left(\partial_l\rho\right)^2 + \partial_l\sigma\partial_l\rho
\right) \left(\sigma
\left(\partial_l^2 \sigma + \partial_l\rho\partial_l
\sigma + (\partial_l\sigma)^2 \right) \right) \nn
&& + {4 \over 3} \e^{-2\sigma}
\left(\partial_l^2 + \partial_l\rho \partial_l
+ \partial_l\sigma\partial_l \right)
\left( \sigma\e^{2\sigma} \left(\partial_l^2 \sigma
+ \partial_l\rho\partial_l \sigma
+ (\partial_l\sigma)^2 \right) \right) \nn
&& - {1 \over 3}\left\{-2\left(\partial_l\sigma\right)^2
\left(\partial_l^2 \rho + \left(\partial_l\rho\right)^2
+ \partial_l\sigma\partial_l\rho \right) \right. \nn
&& \left.\left.
+ \e^{-2\sigma}\left(\partial_l^2 + \partial_l\rho \partial_l
+ \partial_l\sigma\partial_l \right)
\left(\e^{2\sigma}\left(\partial_l\sigma\right)^2 \right)\right\}
\right\} \nn
&& -\left\{b''+ {2 \over 3}(b+b')\right\} r_0^2 \e^{2\rho + 4\sigma}
\nn
&& \times
\left[\e^{-2\sigma}\left(\partial_l^2 + \partial_l\rho \partial_l
+ \partial_l\sigma\partial_l \right) \left( \e^{2\sigma}
\left(\partial_l^2\sigma + \partial_l\rho\partial_l \sigma
+ 2 \left(\partial_l\sigma\right)^2\right) \right) \right. \nn
&& \left. - 3\left(\partial_l^2 \sigma + \partial_l\rho\partial_l \sigma
+ 2 \left(\partial_l\sigma\right)^2\right)^2\right] \\
&& +r_0^2 \e^{2\rho+4 \sigma} \left[ -{4 \over 3}b\left(
\partial_l^2 \rho + \left(\partial_l\rho\right)^2
+ \partial_l\sigma\partial_l\rho \right)^2
\ln \left( {R_4 \over \mu^2 } \right) \right. \nn
&& +{4 \over 3}b\e^{-2\sigma}
\left(\partial_l^2 + \partial_l\rho \partial_l
+ \partial_l\sigma\partial_l \right)
\left\{ \e^{2\sigma}\left(
\partial_l^2 \rho+ \left(\partial_l\rho\right)^2
+ \partial_l\sigma\partial_l\rho \right)
\ln \left( {R_4 \over \mu^2 } \right)\right\} \nn
&& + {4b\e^{-4\sigma} \over 3 r_0^2}
\ln \left({R_4 \over \mu^2 } \right)
-{4 \over r_0^2}\left({b \over 3}+b'\right)\e^{-2\sigma}
\left(\partial_l^2 + \partial_l\rho \partial_l
+ \partial_l\sigma\partial_l \right)
\ln \left({R_4 \over \mu^2 } \right) \nn
&& +{4 \over 3}b \left(\partial_l^2 \rho
+ \left(\partial_l\rho\right)^2
+ \partial_l\sigma\partial_l\rho \right)
\left(\partial_l^2 + \partial_l\rho \partial_l
+ \partial_l\sigma\partial_l \right)
\left\{\ln \left( {R_4 \over \mu^2 } \right)\right\} \nn
&& -{4 \over 3}b\e^{-2\sigma}
\left(\partial_l^2 + \partial_l\rho \partial_l
+ \partial_l\sigma\partial_l \right)
\left\{ \e^{2\sigma} \left(\partial_l^2 + \partial_l\rho \partial_l
+ \partial_l\sigma\partial_l \right)
\left\{\ln \left( {R_4 \over \mu^2 } \right)\right\}\right\} \nn
&& + {4\e^{2\sigma} \left(\partial_l^2 \rho
+ \left(\partial_l\rho\right)^2
+ \partial_l\sigma\partial_l\rho \right) \over R_4 }
\left\{{2 \over 3}b\e^{2\sigma}
\left(\partial_l^2 \rho+ \left(\partial_l\rho\right)^2
+ \partial_l\sigma\partial_l\rho \right)^2 \right. \nn
&& + {2b \e^{-4\sigma} \over 3 r_0^2} + {\e^{-2\sigma} \over 3}b
\left(\partial_l^2 + \partial_l\rho \partial_l
+ \partial_l\sigma\partial_l \right) R_4 \nn
&& \left. - {4 \over r_0^2}\left({b \over 3}+b'\right) \e^{-2\sigma}
\left(\partial_l^2 \rho
+ \left(\partial_l\rho\right)^2
+ \partial_l\sigma\partial_l\rho \right) \rho \right\} \nn
&& + \e^{-2\sigma}\left(\partial_l^2 + \partial_l\rho \partial_l
+ \partial_l\sigma\partial_l \right)
\left\{{2\over R_4 }\left\{{2 \over 3}b \e^{4\sigma}\left(
\partial_l^2 \rho + \left(\partial_l\rho\right)^2
+ \partial_l\sigma\partial_l\rho \right)^2 \right. \right. \nn
&& + {2b \over 3r_0^2}
+ {1 \over 3}b\e^{2\sigma}\left(\partial_l^2 + \partial_l\rho \partial_l
+ \partial_l\sigma\partial_l \right)R_4  \nn
&& \left. \left. \left.
- {4 \e^{2\sigma} \over r_0^2} \left({b \over 3}+b'\right)
\left(\partial_l^2 \rho
+ \left(\partial_l\rho\right)^2
+ \partial_l\sigma\partial_l\rho \right) \right\}\right\}\right] \ , \nn
\label{sigma}
0&=& -{ r_0^2 \e^{2\rho} \over 16\pi G}
\Biggl[-2\e^{4\sigma}\left\{3\partial_l^2 \sigma
+ 4\partial_l\rho\partial_l \sigma
+ 6 \left(\partial_l\sigma\right)^2 + \partial_l^2 \rho
+ \left(\partial_l\rho\right)^2 \right\}
+ {4 \over r_0^2}\e^{2\sigma} \Biggr] \nn
&& + b' r_0^2 \e^{2\rho} \left[ 2 \e^{2\sigma}\left(\partial_l^2
+ \partial_l\rho \partial_l + \partial_l\sigma\partial_l
\right)\left(\e^{2\sigma}\left(\partial_l^2 \sigma
+ \partial_l\rho\partial_l \sigma
+ (\partial_l\sigma)^2 \right) \right)\right. \nn
&& +{4 \over 3}\left( \e^{4\sigma} \left(\partial_l^2 \sigma
+ \partial_l\rho\partial_l \sigma
+ (\partial_l\sigma)^2 \right)
\left(\partial_l^2 \rho + \left(\partial_l\rho\right)^2
+ \partial_l\sigma\partial_l\rho \right) \right. \nn
&& \left. + \e^{2\sigma} \left(\partial_l^2 + \partial_l\rho \partial_l
+ \partial_l\sigma\partial_l \right)
\left(\sigma \e^{2\sigma} \left(\partial_l^2 \rho
+ \left(\partial_l\rho\right)^2 + \partial_l\sigma\partial_l\rho \right)
\rho\right)\right) \nn
&& \left. - {8\e^{2\sigma} \over 3 r_0^2}
\left(\partial_l^2 \sigma+ \partial_l\rho\partial_l \sigma
+ (\partial_l\sigma)^2 \right)
- {\e^{2\sigma} \over 3}\partial_l R_4\partial_l \sigma \right] \nn
&& -\left\{b''+ {2 \over 3}(b+b')\right\} r_0^2 \e^{2\rho+2\sigma} \left[
-{1 \over 2}\left(\partial_l^2 + \partial_l\rho \partial_l
+ \partial_l\sigma\partial_l \right) R_4
+ \partial_l R_4\partial_l \sigma \right. \\
&& + 3 \left(\partial_l^2 + \partial_l\rho \partial_l
+ \partial_l\sigma\partial_l \right) \left\{ \e^{2\sigma}
\left(\partial_l^2\sigma + \partial_l\rho\partial_l \sigma
+ 2 \left(\partial_l\sigma\right)^2\right)\right\} \nn
&& \left. -6 \e^{-\sigma -\rho} \partial_l\left\{ \e^{\rho + 3\sigma}
\partial_l\sigma\left(\partial_l^2 \sigma
+ \partial_l\rho\partial_l \sigma + 2 \left(\partial_l\sigma\right)^2
\right)\right\}  \right] \ . \nonumber
\eea
The equations (\ref{cons}) and (\ref{rho}) contain the 6th order
derivatives of $\rho$, $\partial_l^6\rho$ and the equations (\ref{cons}),
(\ref{rho}) and (\ref{sigma}) contain the 4th order derivatives of
$\sigma$.
Therefore we need to impose the
initial conditions including 5th order derivative of $\rho$ and 3rd
order derivative of $\sigma$:
\bea
\label{inicon}
&& \partial_l\rho |_{l=0}=\partial_l^2\rho |_{l=0}
=\partial_l^3\rho |_{l=0}=\partial_l^4\rho |_{l=0}
=\partial_l^5\rho |_{l=0}=0 \nn
&& \partial_l\sigma |_{l=0}=\partial_l^2\sigma |_{l=0}
=\partial_l^3\sigma |_{l=0}=0\ .
\eea
The equations (\ref{cons}) and (\ref{rho}) contain
the 5th order derivatives of $\sigma$ from $\partial_l^4R_4$ but
$\partial_l^5 \sigma$ always appears in the form
$\partial_l^5 \sigma \partial \rho$. Therefore these terms vanish
at $l=0$. This tells that we cannot impose the initial condition
$\partial_l^4\sigma=0$ since we cannot solve the differential
equations with respect to $\partial_l^5 \sigma$ at $l=0$.

We should also note that all the equations (\ref{cons}),
(\ref{rho}) and (\ref{sigma}) are not the dynamical equations of
motion but the combinations with the constraint.
The constraint can be found by putting $l=0$ under the initial
conditions (\ref{inicon}), when the equations (\ref{cons}) and
(\ref{rho}) have the following form:
\begin{eqnarray}
\label{eq1}
0&=& \left( b+ b'+ \frac{3}{2}b''
+ 8b'\sigma_0 \right)\,\sigma^{(4)}_0 +
14b{\e^{-2\,\sigma{}_0}}{{r_{0}}^2}\,\rho^{(6)}_0\, ,\\
0&=&\frac{3{\e^{-2\sigma{}_0}}}{8\,G\,\pi}
- 2b{\e^{-4\sigma{}_0}}\ln \left(
{\frac{2}{{\mu^2}{{r_{0}}^2}}}\right) \nonumber \\
\label{eq2}
&& + {{r_{0}}^2} \left( b + b' + \frac{3}{2}b''
- 2b'\sigma{}_0 \right)\, \sigma^{(4)}_0
-  b{\e^{-2\sigma{}_0}}{{r_{0}}^4}\, \rho^{(6)}_0\, ,\\
\label{eq3}
0&=&\frac{{\e^{- 2\,\sigma{}_0}}}{G\,\pi \,{{r_{0}}^2}} +
4  \left( 2\,b + 3\,b'' \right)\,\sigma^{(4)}_0 \, .
\end{eqnarray}
Equations (\ref{eq1}), (\ref{eq2}) and (\ref{eq3}) will be compatible
if $\sigma_0$ satisfies to the following equation:
\begin{eqnarray}
0&=& -224\,b\,\left( 2\,b + 3\,b'' \right) \,G\,\pi \,
   \ln \left({\frac{2}{{\mu^2}\,{{r_{0}}^2}}}\right)
\nonumber \\
\label{eq4}
&& +   {\e^{2\,\sigma{}_0}}\,\left( 54\,b - 30\,b' +
     81\,b'' + 40\,b'\,\sigma{}_0 \right) \, .
\end{eqnarray}
Then we can choose the initial conditions as (\ref{inicon}) and
(\ref{eq4}) with arbitrary $\rho |_{l=0}$. Then under these initial
conditions, we can use any two of equations (\ref{cons}),
(\ref{rho}) and (\ref{sigma}) as dynamical equations of motion.
Equation (\ref{eq4}) admits the real solution for $\sigma_0$ if
\be
\label{eq5}
b\,\left( 112\,b + 168\,b'' \right) \,
     {\e^{-{\frac{1}{2}} + {\frac{27\,\left( 2\,b + 3\,b'' \right) }
           {20\,b'}}}}\,G\,\pi \,
     \ln \left({\frac{2}{\mu^2\,{{{r_0}}^2}}}\right) < -{5\,b'}
\ee
and from (\ref{eq4}) we have the restriction
\be
\sigma_0 \ge \frac{1}{4}+\frac{54\,b+81\,b''}{40\,|b'|}
\ee
 From equations (\ref{cons}), (\ref{rho}), (\ref{sigma}) and initial
conditions (\ref{inicon}) we get the following
behavior of $\rho$ and
$\sigma$ near $l=0$
\bea
\label{eq6}
\rho(l)&=&\rho_0+\frac{2 b+2 b'+3 b''+16 b' \sigma_0}{80640\,
G\,\pi\, r_0{}^4\, b\, (2b+3b'')
}\,l^6+ O(l^8)
\\
\label{eq7}
\sigma(l)&=&\sigma_0 -\frac{\e^{-2\sigma_0}}{96\,G\,\pi\,
r_0{}^2\,(2b+3b'')}\, l^4  +O(l^6)
\eea
Then for initial conditions (\ref{inicon})
$\sigma(l)$ and $\rho(l)$ are decreasing functions near $l=0$.

For the numerical calculation as an example
we will consider ${\cal N}=4$  $SU(N)$ super-YM
theory where (see, for example, \cite{BO})
\be
\label{eq8}
b=-b'=\frac{N^2-1}{(8\,\pi)^2},
\qquad
b''=0.
\ee
Such theory became popular recently in relation with AdS/CFT 
correspondence. We take it as typical example of GUT.
Note also that as it was mentioned in ref.\cite{BO}
the explicit choice of $b''$ does not influence the equations 
of motion.

Then from (\ref{eq6}-\ref{eq7}) we get
\bea
\label{eq9}
\rho(l)&=&\rho_0+\frac{-\sigma_0}{10080\,
G\,\pi\, r_0{}^4\, b
}\,l^6+ O(l^8),
\\
\label{eq10}
\sigma(l)&=&\sigma_0 -\frac{\e^{-2\sigma_0}}{198\,G\,\pi\,
r_0{}^2\,b}\, l^4  +O(l^6)
\eea
and from (\ref{eq4}-\ref{eq5}) we obtain
%\be
%\label{eq11}
%112\,b\,G\,\pi \,\ln (\frac{2}{\mu^2\,{r_0}^2}) < 5\,\e^{16/5}
%\ee
\bea
\label{eq12}
(\mu r_0)^2 > 2 & \to & b>0,
\quad
\sigma_0 > 21/10
\\
\label{eq13}
(\mu r_0)^2 = 2 & \to & b>0,
\quad
\sigma_0=21/10
\\
\label{eq14}
(\mu r_0)^2 < 2 & \to & 0<b<\frac{5\,\e^{16/5}}{112\,G\,\pi \,
\ln (\frac{2}{\mu^2\,{r_0}^2})},
\quad
\frac{8}{5}<\sigma_0<\frac{21}{10}.
\eea
Let us choose for numeric investigation, that
$$
\label{eq15}
G=1, \quad \mu^2=2, \quad {r_0}^2=2/\mu^2=1.
$$
The results of numerical calculation are given by
Figures 1 and 2 for redshift and shape functions
where $N=5$.

\epsffile[20 280 500 450]{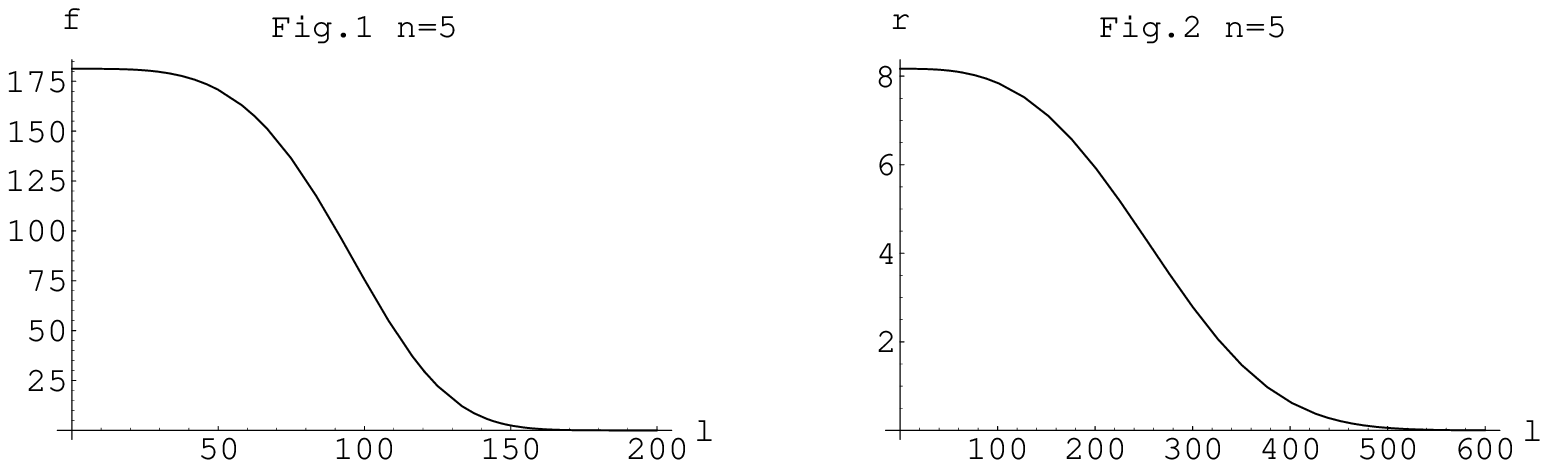}

\noindent
Hence we showed that for above choice of initial
conditions the inducing of primordial wormholes
at the early Universe is rather unrealistic. The throat radius
of wormhole quickly shrinks to zero.

Let us now consider another choice of initial conditions
\bea
\label{inicon2}
&& \partial_l\rho |_{l=0}=\partial_l^2\rho |_{l=0}
=\partial_l^3\rho |_{l=0}=0,
\quad
\partial_l^4\rho |_{l=0} \ne 0  \nn
&& \partial_l\sigma |_{l=0}=\partial_l^2\sigma |_{l=0}
=\partial_l^3\sigma |_{l=0}=0\ .
\eea
 From equations (\ref{cons}), (\ref{rho}), (\ref{sigma}) and initial
conditions (\ref{inicon2}) we get the following behavior of $\rho$ and
$\sigma$ near $l=0$
$$
\partial_l^4\sigma |_{l=0}  =
\frac{1}{792{}{b^2}{}G{}\pi }
\biggl[
{{-99{}b}{{\e^{-2{}\sigma_0}}}} +
     4{}{b^2}{}G{}\pi {}\rho_0{}\sigma_0{}
      \left( -51 + 36{}{\e^{4{}\sigma_0}} +
        20{}{\e^{4{}\sigma_0}}{}\rho_0{}{{\sigma_0}^2}
         \right)
$$
\be
- 4\varepsilon{}{\sqrt{\pi }}{}\rho_0{}\sigma_0{}
      {\sqrt{{b^3}{}G{}\left( 99{}{\e^{2{}\sigma_0}}{}
             \left(21 - 10{}\sigma_0 \right)  +
            b{}G{}\pi {}{{\left( -51 + 36{}{\e^{4{}\sigma_0}} +
                  20{}{\e^{4{}\sigma_0}}{}\rho_0{}
                   {{\sigma_0}^2} \right) }^2} \right) }}
\, \biggr]
\ee
$$
\partial_l^4\rho |_{l=0}=
\frac{1}{132\,{b^2}\,G\,{ \sqrt{\pi} }}
\biggl[
{b^2}\,G\,\sqrt{\pi}\,\left( 51 - 36\,{\e^{4\,\sigma_0}} -
          20\,{\e^{4\,\sigma_0}}\,\rho_0\,{{\sigma_0}^2}
            \right)
$$
\be
+  \varepsilon{\sqrt{{b^3}\,G\,
            \left( 99\,{\e^{2\,\sigma_0}}\,
               \left( 21 - 10\,\sigma_0 \right)  +
              b\,G\,\pi \,{{\left( -51 + 36\,{\e^{4\,\sigma_0}} +
                    20\,{\e^{4\,\sigma_0}}\,\rho_0\,
                     {{\sigma_0}^2} \right) }^2} \right)
}}\, \biggr]
\ee
where $\varepsilon=\pm1$.

Moreover we obtain the conditions
\be
\partial_l^5\rho |_{l=0}=\partial_l^5\sigma |_{l=0}= 0,
\ee
\centerline{or}
\be
891\,{\e^{2\,\sigma_0}}\,\left(10\,\sigma_0 - 21 \right)  =
  8\,b\,G\,\pi \,{{\left( -51 + 36\,{\e^{4\,\sigma_0}} +
        20\,{\e^{4\,\sigma_0}}\,\rho_0\,{{\sigma_0}^2}
         \right) }^2},
\quad
\sigma_0>\frac{21}{10}.
\ee
For case $\sigma_0=21/10$ we have
\bea
\partial_l^4\sigma |_{l=0}
&=&
\frac{1}{2200}\left(
\frac{-275\,\e^{-21/5}}{b\,G\,\pi} +
         (1680\,\e^{42/5}- 2380)\,\rho_0 +
         4116\,\e^{42/5}\,{\rho_0}^2
\right),
\\
\partial_l^4\rho |_{l=0}
&=&\frac{1}{110}
\left( 85 - 60\,\e^{42/5} - 147\,\e^{42/5}\,\rho_0 \right)
\eea
Then functions $\sigma(l)$ and $\rho(l)$ will increase near $l=0$
if
$$
\rho_0<
\frac{5}{294\,b\,G}\,
\left[ b\,G\,\left( -12 + 17\,\e^{-42/5}\right)
         - \frac{\e^{-42/5}}{{\sqrt{\pi }}}{\sqrt{b\,G\,
              \left( 231\,{\e^{{\frac{21}{5}}}} +
                b\,{{\left( 17 - 12\,{\e^{{\frac{42}{5}}}} \right) }^2}\,G\,
                 \pi  \right) }}
\,\right]
$$
for $G=1$ and $b=(N^2-1)/(8\pi)^2$ we have $\rho_0<-0.4081$.
The explicit numerical study (see Figures 3 and 4) shows
that shape and redshift functions are increasing
giving the window for inducing of primordial
wormholes.

\leavevmode
\epsffile[15 265 463 396]{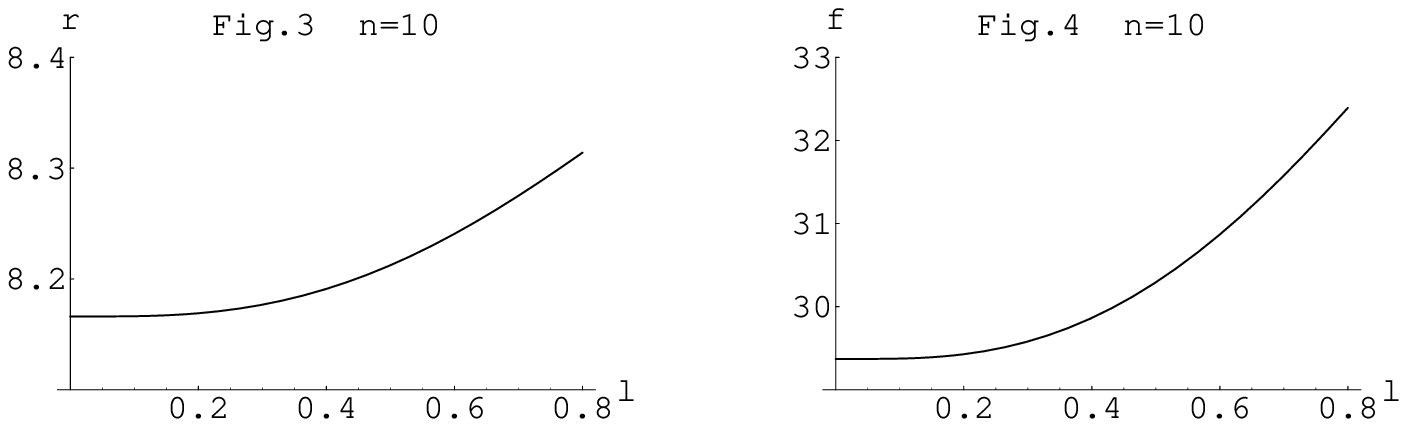}

\noindent
On these figures $\sigma_0=21/10$ and $\rho_0=-0.41$.

Hence we numerically proved that at least for some
initial conditions GUTs at the early Universe may
help in producing of primordial wormholes. It is of course the
open question which initial conditions from the two
classes presented above lead to more stable configuration.
Moreover, due to complicated structure of field equations
and initial conditions itself we cannot classify
from the very beginning the initial conditions as supporting (or not)
wormholes production. Nevertheless for any specific GUT
under discussion above study may be easily repeated
and principal possibility of wormholes inducing may be shown 
at least numerically.

\bigskip

\noindent
{\bf Acknowledgments}. The research by
SDO was partially supported by a RFBR Grant N\,99-02-16617,
by Saxonian Min. of Science and Arts and by Graduate College
``Quantum Field Theory" at Leipzig University. The research by
KEO was partially supported by a RFBR Grant N\,99-01-00912 and 
the research by OO has been partially supported by 
CONACyT grants 3898P-E9608 and 28454-E.


\begin{thebibliography}{99}
\bibitem{12}  M.S. Morris and K.S. Thorne, Am.J.Phys.
\underline{56} (1988) 395;
M.S. Thorne, K.S. Thorne and U. Yurtsever,
Phys.Rev.Lett. \underline{61} (1988) 1446
\bibitem{13}
G. Gibbons, preprint hep-th/9801106.
\bibitem{15}
D.Hochberg, A. Popov and S.N. Syshkov,
{\it Phys.Rev.Lett.} {\bf 78} (1997) 2050.
\bibitem{16} V. Khatsymovsky, In proceedings of II Int.Conf.
Quantum Field Theory and Gravity, 
Eds.I.L. Buchbinder and K.E. Osetrin,
TGPU Publishing, Tomsk, 1997.
\bibitem{NOOO}
S. Nojiri, O. Obregon, S.D. Odintsov and K.E. Osetrin,
{\it Phys.Lett.} {\bf B449} (1999) 173.
\bibitem{BOS} I.L. Buchbinder, S.D. Odintsov and I.L. Shapiro,
{\sl Effective Action in Quantum Gravity}, IOP Publishing,
Bristol and Philadelphia, 1992.
\bibitem{NO} S. Nojiri and S.D. Odintsov, {\it Phys.Rev.}  {\bf D59}
(1999) 044026.
\bibitem{R} R.J. Reigert, {\it Phys.Lett.} {\bf B134} (1984)56;
E.S. Fradkin and A. Tseytlin, {\it Phys.Lett.} {\bf B134} (1984) 187;
I.L. Buchbinder, S.D. Odintsov and I.L. Shapiro, {\it Phys.Lett.}
{\bf B162} (1985) 92, 
for a recent review see D. Anselmi, hep-th/9903059.
\bibitem{NOa} S. Nojiri and S.D. Odintsov, hep-th/9802160,
{\it Int.J.Mod.Phys.} {\bf A}, to appear.
\bibitem{BO} I. Brevik and S.D. Odintsov, hep-th/9902184, 
{\it Phys.Lett.} {\bf B}, to appear.
\bibitem{AMO} I. Antoniadis and E. Mottola, {\it Phys.Rev.}  {\bf D45}
(1992) 2013; S.D. Odintsov, {\it Z.Phys.} {\bf C54} (1992) 531.
\end{thebibliography}
\end{document}